\documentclass{aa}
\usepackage[colorlinks,citecolor=blue,linkcolor=blue,urlcolor=blue]{hyperref}
\title{Orientation of the crescent image of M~87*}
\titlerunning{Orientation of crescent image of M~87*}
\author{
Krzysztof~Nalewajko\thanks{\email{knalew@camk.edu.pl}},
Marek~Sikora
and Agata~R{\'o}{\.z}a{\'n}ska
}
\authorrunning{Nalewajko {et~al.}}
\institute{
Nicolaus Copernicus Astronomical Center, Polish Academy of Sciences, Bartycka 18, 00-716 Warsaw, Poland
\label{inst_ncac}
}

\abstract
{The first image of the black hole (BH) M 87* obtained by the Event Horizon Telescope (EHT) has the shape of a crescent extending from the E to WSW position angles, while the observed direction of the large-scale jet is WNW.
Images based on numerical simulations of BH accretion flows suggest that on average the projected BH spin axis should be oriented SSW.
We explore highly simplified toy models for geometric distribution and kinematics of emitting regions in the Kerr metric, perform ray tracing to calculate the corresponding images, and simulate their observation by the EHT to calculate the corresponding visibilities and closure phases.
We strictly assume that (1) the BH spin vector is fixed to the jet axis, (2) the emitting regions are stationary and symmetric with respect to the BH spin, and that (3) the emissivities are isotropic in the local rest frames.
Emission from the crescent sector between SSE and WSW can be readily explained in terms of an equatorial ring with either circular or plunging geodesic flows, regardless of the value of BH spin.
In the case of plane-symmetric polar caps with plunging geodesic flows, the dominant image of the cap located behind the BH is sensitive to the angular momentum of the emitter.
Within the constraints of our model, we have not found a viable explanation for the observed brightness of the ESE sector.
Most likely, the ESE `hotspot' has been produced by a non-stationary localised perturbation in the inner accretion flow.
Alternatively, it could result from locally anisotropic synchrotron emissivities.
Multi-epoch and polarimetric results from the EHT will be essential to verify the theoretically expected alignment of the BH spin with the large-scale jet.}

\keywords{
Black hole physics
--- Galaxies: active
--- Galaxies: individual: M87
--- Gravitation
--- Relativistic processes
}

\begin{document}

\maketitle

\section{Introduction}

The EHT Collaboration presented the first resolved image of a photon ring surrounding the supermassive BH dubbed M87* \citep{2019ApJ...875L...1E}. The image obtained at the wavelength of $\lambda = 1.3\,{\rm mm}$ and the angular resolution of $\simeq 20\;{\rm \mu as}$ can be best described as an asymmetric crescent of diameter $\simeq 42\;{\rm \mu as}$ with brightness enhanced on the southern side \citep{2019ApJ...875L...6E}, more precisely for position angles in the range ${\rm PA} \in [70^\circ:270^\circ]$ (NEE -- W) \citep[Fig. 27]{2019ApJ...875L...4E}. On the other hand, interferometric observations at longer wavelengths reveal a well-known relativistic jet: at $3.5\;{\rm mm}$ wavelength the jet collimation and acceleration zone can be probed up to $\sim 1\;{\rm mas}$ \citep{2016ApJ...817..131H,2018A&A...616A.188K}; at $7\;{\rm mm}$ wavelength mildly superluminal motions can be traced up to $\simeq 6\;{\rm mas}$ \citep{2016A&A...595A..54M,2018ApJ...855..128W}. The jet is directed at the mean position angle of ${\rm PA}_{\rm jet} \simeq 288^\circ$ (NWW), and this projected direction has been stable over decades of observations.

The EHT image of M87* has been compared with ray-traced images obtained from an extensive library of general relativistic magnetohydrodynamical (GRMHD) simulations of geometrically thick accretion onto BHs of various spins \citep{2019ApJ...875L...5E}.
The enhanced brightness on the southern side of the ring has been interpreted as resulting from the Doppler beaming of mildly relativistic plasma flow that is approaching the observer.
A sense of rotation for the BH (rather than for the large-scale accretion flow) can be deduced from this as clockwise (CW; corresponding to the observer inclination angle with respect to the BH spin $i > 90^\circ$).
However, the projected orientations of the BH spin vectors --- obtained by rotating the simulated images to minimise the difference from the observed image --- were found to span a range of ${\rm PA}_{\rm BH} \simeq 150^\circ - 280^\circ$, with the mean value of $\simeq 205^\circ$ and standard deviation of $\simeq 55^\circ$ (SSE -- W; top panels of Fig. 9 in \citealt{2019ApJ...875L...5E}).
The mean offset from ${\rm PA}_{\rm jet}$ is $\simeq 80^\circ$ clockwise (CW), which is equivalent to $\sim 1.5\sigma$.
The predictions of GRMHD simulations specifically tailored to the case of M87 are that the major axis of the bright part of the ring should be oriented parallel to the projected jet axis \citep{2012MNRAS.421.1517D,2016A&A...586A..38M,2019A&A...632A...2D}.

There are strong arguments to support the assumption that relativistic jets should be directed along the BH spin axis --- the leading scenario for launching jets is the Blandford-Znajek mechanism \citep{1977MNRAS.179..433B}, numerical simulations demonstrated the robustness of jet alignment even for flipping BH spins \citep{2013Sci...339...49M}, and there is no evidence for jet precession in the particular case of M87 \citep[see Section 7.3.(2) in][]{2019ApJ...875L...5E}.

In the GRMHD simulations of accretion-jet systems \citep{2019ApJ...875L...5E}, the synchrotron emission is produced mainly in the accreting torus or in the outer jet funnel, but not in the polar jet regions.
Emission from the polar regions requires additional physical mechanisms of energy dissipation beyond the regime of ideal magnetohydrodynamics (MHD).
In particular, these regions could be related to particle-starved gaps in the BH magnetospheres suggested as the sites of very-high-energy gamma ray emission
(e.g., \citealt{2011ApJ...730..123L}, \citealt{2015ApJ...809...97B}, \citealt{2018ApJ...852..112K})
that has actually been observed in M87 and found to be variable on the time scale of days \citep[e.g.,][]{2006Sci...314.1424A, 2012ApJ...746..151A}.
The polar regions could also be related to the lamppost model for the origin of hard X-ray emission in AGNs and X-ray binaries \citep[e.g.,][]{1996MNRAS.282L..53M, 2004MNRAS.349.1435M}.
The first kinetic simulations of BH magnetospheres with pair creation demonstrate the formation of charge gaps at intermediate latitudes for low enough rate of particle supply \citep{2019PhRvL.122c5101P}.
For these reasons, we find it particularly important to include in our study a simplified case of polar caps.

In this letter, we consider highly simplified (stationary, axisymmetric) toy models for the distribution of emitting regions around the BH M87*, and we ray-trace the corresponding images in the Kerr metric in the limit of optically thin emission with uniform fluid-frame emissivity.
We fix the BH spin axis at position angle of ${\rm PA}_{\rm jet} = 288^\circ$ and inclination of $i = 162^\circ$. We assume that the emission process is isotropic in local rest frames, and we consider the effect of Doppler beaming due to azimuthal and radial motions of the emitting fluid.
We find that our model can naturally explain a section of the crescent image that extends between SSE and SWW position angles, however, it cannot explain the emission observed in the SEE sector.

\section{Methods}

We perform ray tracing of scalar radiation in the Kerr metric for BH mass $M$ and spin $a$ ($G = 1 = c$),
\begin{eqnarray}
{\rm d}s^2 &=&
-\left(1-\frac{2Mr}{\Sigma}\right){\rm d}t^2
+\frac{\Sigma}{\Delta}\,{\rm d}r^2
+\Sigma\,{\rm d}\theta^2
\\&&\nonumber
+\left(r^2+a^2+\frac{2Mra^2\sin^2\theta}{\Sigma}\right)\sin^2\theta\,{\rm d}\phi^2
\\&&\nonumber
-\frac{4Mra\sin^2\theta}{\Sigma}{\rm d}t\,{\rm d}\phi\,,
\\
\Sigma &=& r^2 + a^2\cos^2\theta\,,
\quad
\Delta = r^2 + a^2 - 2Mr\,,
\nonumber
\end{eqnarray}
using the Boyer-Lindquist coordinates $x^\mu = (t,r,\theta,\phi)$, employing a newly developed numerical code.
We integrate the null geodesics {$x^\mu(\lambda)$ over the affine parameter ${\rm d}\lambda = {\rm d}t / p^t$, with the 4-momentum $p^\mu = {\rm d}x^\mu / {\rm d}\lambda$, starting from a stationary observer of position $x_{\rm obs}^\mu = (t_{\rm obs},r_{\rm obs} = 100 M,\theta_{\rm obs} = i,\phi_{\rm obs} = 0)$ and 4-velocity $u_{\rm obs}^\mu = u_{\rm obs}^t(1,0,0,0)$.}
We perform first-order explicit integration {backward in time}, with an adaptive time step ${\rm d}t \propto r$.
The integration involves 4 constants: asymptotic energy $E$, asymptotic angular momentum $L$, Carter constant $Q$, {and square of the 4-momentum $p_\mu p^\mu = 0$} \citep[e.g.,][]{1972ApJ...178..347B,Sikora1979}.
Integration is interrupted either when the photon falls onto the BH ($p^t \ge 10^3$) or when the photon escapes ($r > r_{\rm obs}$).

The emitting regions are assumed to be optically thin with uniform isotropic emissivities with power-law spectrum {$j_{\rm em}(\nu_{\rm em}) \propto \nu_{\rm em}^{-\alpha}$ of spectral index $\alpha$ in their local rest frames.}
{We adopt a spectral index $\alpha = 0.5$, and we checked that varying $\alpha$ has little effect on the resulting image.}
The emission mechanism is not specified, and no absorption of radiation is considered.
{The emitters are allowed to be in motion relative to the local stationary observers with radial velocities $\beta_r = {\rm d}r/{\rm d}t$ and angular velocities $\Omega = {\rm d}\phi/{\rm d}t$, hence their 4-velocities are $u_{\rm em}^\mu = u_{\rm em}^t(1,\beta_r,0,\Omega)$.}
From the covariant radiative transfer equation ${\rm d}_\lambda(I_\nu/\nu^3) = j_\nu/\nu^2$ \citep[e.g.,][]{2018MNRAS.475...43M},
for every geodesic $\mathcal{G}$ that intersects an emitting region $\mathcal{S}$, the observed intensity is integrated as being proportional {to $I_{\rm \nu,obs} \propto \int_{\mathcal{G}\cap\mathcal{S}} {\rm d}\lambda\,g^{2+\alpha}$, where $g = [p_\mu(\lambda_{\rm obs}) u_{\rm obs}^\mu] / [p_\mu(\lambda_{\rm em}) u_{\rm em}^\mu]$ is the factor that combines gravitational redshift with the Doppler boost due to the motion of the emitter.}
The obtained images are scaled in the units of $\theta_{\rm g} = \arcsin(M/r_{\rm obs})$, Gaussian-smoothed {(using the {\tt scipy.ndimage.gaussian\_filter} tool)} with the angular resolution of $5.3\theta_{\rm g} \simeq 20\;{\rm\mu as}$, and normalised to the peak flux density of the EHT image.
The centring of the EHT image relative to the simulated images is approximate.

We consider very simple geometries for stationary emitting regions that are axisymmetric with respect to the BH spin axis and plane-symmetric with respect to the BH equatorial plane. As such, they can be defined by the location of their boundaries in the coordinates $(r,\theta)$ (Figure \ref{fig_geometry}). In particular, we consider regions defined by $r \le r_{\rm max}$ and $0 \le \theta_{\rm min} \le \theta \le \theta_{\rm max} \le \pi/2$, and for plane symmetry we also include the mirror regions that satisfy $\theta_{\rm min} \le (\pi-\theta) \le \theta_{\rm max}$.
In our calculations we adopt an outer radius $r_{\rm max} = 6 R_{\rm g}$, since lower values of $r_{\rm max}$ result in images that are too compact.

\subsection{Simulating EHT observations}

We performed simulated EHT observations of our selected model images with the {\tt eht-imaging} software package (version 1.1.1) \citep{2016ApJ...829...11C,2018ApJ...857...23C}.
For the {\tt Image} class object, we adopted the following parameters: the sky coordinates of M87* ($\alpha = 12^h 30^m 49^s.4234$, $\delta = 12^\circ 23^m 28^s.0439$), the observation date ${\rm MJD = 57854}$ (2017 April 11th), and the observational frequency $\nu_{\rm obs} = 229.1\;{\rm GHz}$.
The simulated observations (the {\tt Image.observe} method) were performed with the amplitude and phase calibrations, and with the following parameters: the scan integration time equal to the the time separation between scans ${\tt tint\_sec} = {\tt tadv\_sec} = 5\,{\rm min}$, and the bandwidth $\Delta\nu_{\rm obs} = 2\;{\rm GHz}$.

For the calculation of closure phases (cf. Section 2.1.2 in \citealt{2019ApJ...875L...4E}), we consider 10 distinct telescope triangles, excluding the APEX node and treating the JCMT node as equivalent to the SMA node (excluding also the JCMT -- SMA baseline). The telescopes are ordered in the following sequence: (1) ALMA, (2) SMT, (3) JCMT, (4) LMT, (5) PV, (6) SMA. For every telescope triangle with ordered nodes $(i < j < k)$, the closure phase is calculated as $\Psi_{C,ijk} = s_{ijk}(\Psi_{ik} - \Psi_{ij} - \Psi_{jk})$, where $\Psi_{ij}$ are the visibility phases measured for particular ordered baselines $(i < j)$,\footnote{For the simulated observations reported with inverted baselines $(i > j)$, we reverse the signs of $u_{ji} = -u_{ij}$, $v_{ji} = -v_{ij}$ and $\Psi_{ji} = -\Psi_{ij}$.} and $s_{ijk} = \pm 1$ is the \emph{triangle chirality} that depends on whether the sequence $(i,j,k)$ of triangle nodes, as seen from the perspective of M87*, would be visited in clockwise ($s_{ijk} = 1$) or counterclockwise ($s_{ijk} = -1$) order.\footnote{More strictly, we define the triangle chirality as $s_{ijk} = {\rm sign}[(u,v)_{ij} \wedge (u,v)_{jk}]$. The triangle (1,2,4) or ALMA-SMT-LMT changes its chirality from $s_{124} = +1$ to $s_{124} = -1$ at ${\rm UTC} \simeq 5^h.2$. To avoid flipping the sign of $\Psi_{C,124}$, we adopted a fixed value of $s_{124} = +1$.}
To verify our procedure, we simulated EHT observations of the average reconstructed EHT image of M87*, reproducing the observed closure phases with good accuracy.
We also simulated EHT observations of basic crescent images in order to understand how the closure phases respond to varying the crescent orientation.

The results of simulated EHT observations were compared with the actual calibrated photometric EHT data on M87* from the EHT Science Release 1 (SR1) package \citep{SR1}. We consider only the results obtained in the higher-frequency band (centred at $229.1\;{\rm GHz}$), since the results obtained in the lower-frequency band (centred at $227.1\;{\rm GHz}$) are essentially the same for our purposes.
Individual measurements are grouped into scans, within which the time separations between consecutive measurements are $< 36\;{\rm s}$. For each scan, we calculate the mean values of baselines ($\left<u\right>$, $\left<v\right>$) and complex visibility $\left<I\right> = \left<I\right>_{\rm amp}\exp(i\left<\Psi\right>)$, from which the mean values of visibility amplitude $\left<I\right>_{\rm amp}$ and visibility phase $\left<\Psi\right>$ are derived.
We attempted to evaluate the signal-to-noise values for the photometric scans, we noticed however that the {\tt Isigma} values reported in the SR1 dataset appear to be underestimated.
Instead, for each scan we evaluated the measurement uncertainties as $\sigma_I = {\rm rms}({\tt Isigma})$, and for the calculation of significant closure phases we require that $\min_{ij}\{\left<I\right>_{{\rm amp},ij} / \sigma_{I,ij}\} > 0.15$.

\begin{figure}
\includegraphics[width=\columnwidth]{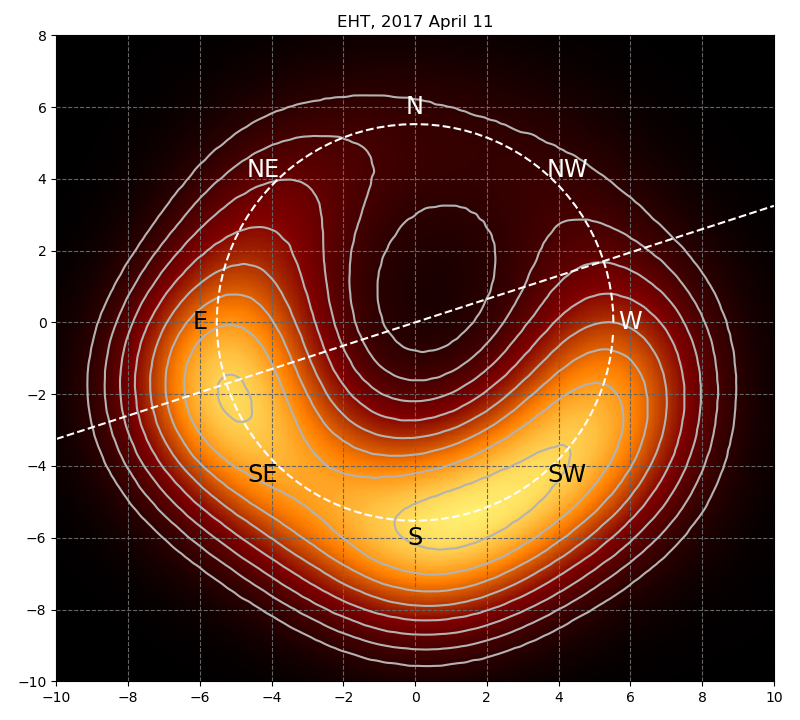}
\caption{Image of BH M87* at 1.3 mm wavelength obtained on April 11, 2017 by the Event Horizon Telescope (adapted from Fig. 3 in \citealt{2019ApJ...875L...1E}). The grey contour levels correspond to brightness temperature values of $(2,2.5,3,...,5.5)\times 10^9\;{\rm K}$. The white dashed lines indicate a photon ring of diameter $42\;{\rm \mu as}$ equivalent to $\simeq 11M$, and the position angle of large-scale jet ${\rm PA} = 288^\circ$ (NWW). The units of the grid are the angular size of the M87* gravitational radius $\theta_{\rm g} \simeq 3.8\;{\rm \mu as}$.}
\label{fig_eht_calib}
\end{figure}

\begin{figure}
\centering
\includegraphics[width=0.9\columnwidth]{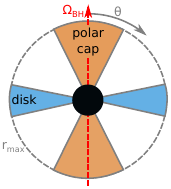}
\caption{Schematic geometry of the axially symmetric emitting regions in the space of Boyer-Lindquist coordinates $(r,\theta)$. The emitting regions are limited to $r < r_{\rm max} = 6M$ and $\theta_{\rm min} < \theta,\pi-\theta < \theta_{\rm max}$. In the case of polar caps (TH1), we adopt $\theta_{\rm min} = 0$ and $\theta_{\rm max} = 30^\circ$. In the case of equatorial disk (TH4), we adopt $\theta_{\rm min} = 75^\circ$ and $\theta_{\rm max} = 90^\circ$.}
\label{fig_geometry}
\end{figure}

\begin{figure*}
\includegraphics[width=\textwidth]{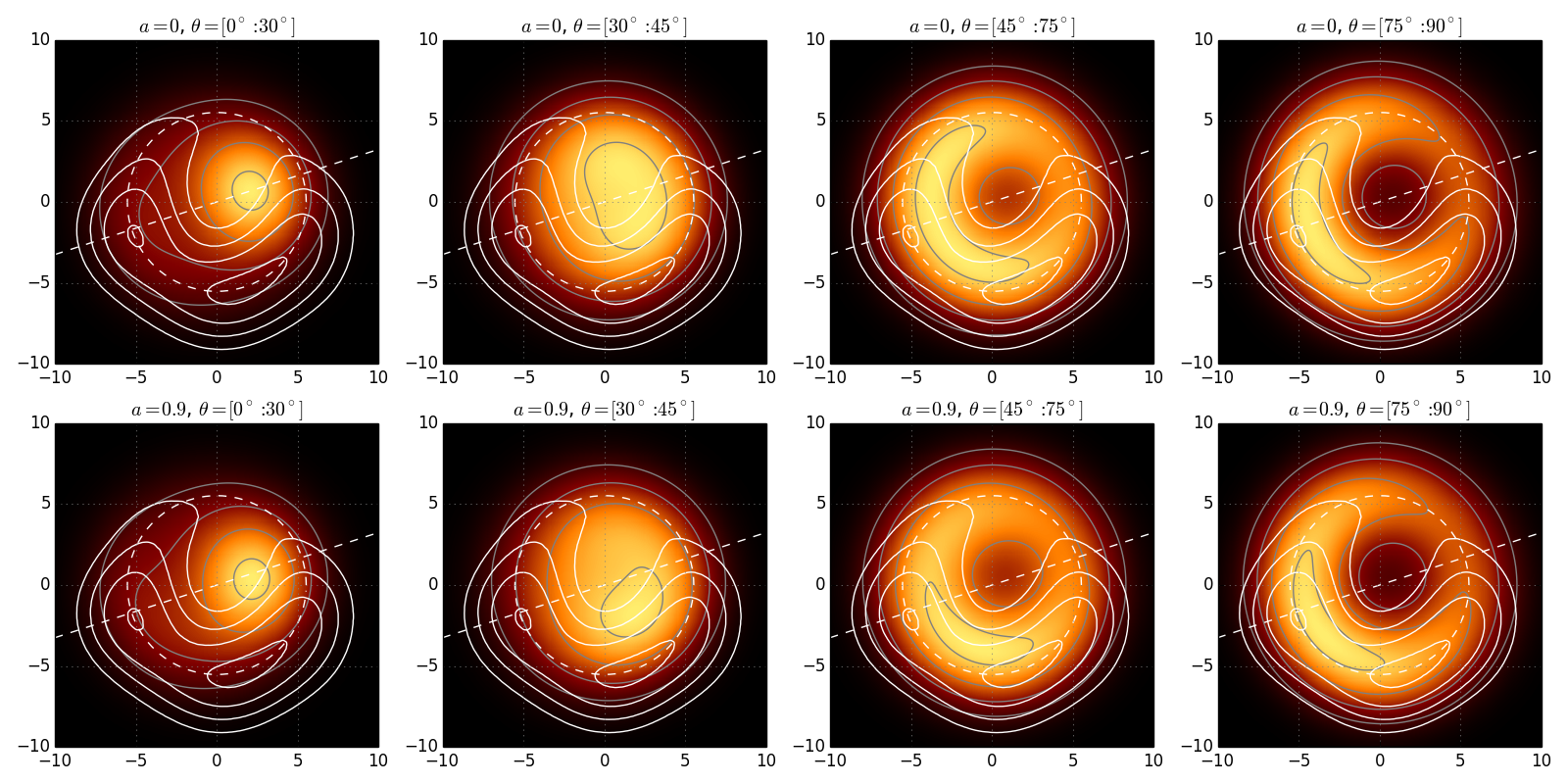}
\caption{A series of simulated images for source regions limited to the radial coordinate values $r \le r_{\rm max} = 6M$ and to different values of the polar coordinate (models TH1-4, from left to right) for two values of BH spin. In these cases, the local emitter is assumed to be stationary ($\beta_r = 0 = \Omega$), and the spectral index is $\alpha = 0.5$. The simulated images are shown with colour shading and grey contours, while the EHT image of M87* is shown with solid white contours. The contour levels correspond to brightness temperature values of $(2.5,3.5,4.5,5.5)\times 10^9\;{\rm K}$, both for the simulated images and for the EHT image. The dashed white circle indicates a photon ring of radius $21\;{\rm \mu as} \simeq 5.5 \theta_{\rm g}$ (best-fit estimate by EHT).
The dashed white line indicates the position angle ${\rm PA}_{\rm jet} = 288^\circ$ of the large-scale jet.}
\label{fig_eht_beta0_spin_theta}
\end{figure*}

\begin{figure*}
\includegraphics[width=\textwidth]{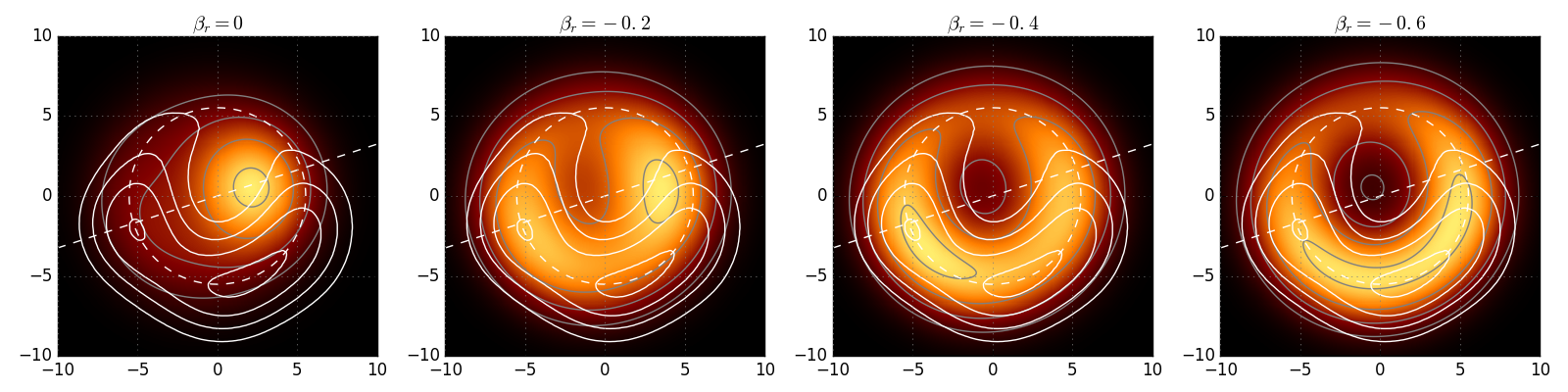}
\caption{The case of polar caps (TH1) for $a = 0.5$, showing the effect of Doppler beaming due to radial infall of the emitting fluid with fixed radial velocity $\beta_r$. See Fig. \ref{fig_eht_beta0_spin_theta} for more description.}
\label{fig_eht_th1_betar}
\end{figure*}

\begin{figure*}
\centering
\includegraphics[width=\textwidth]{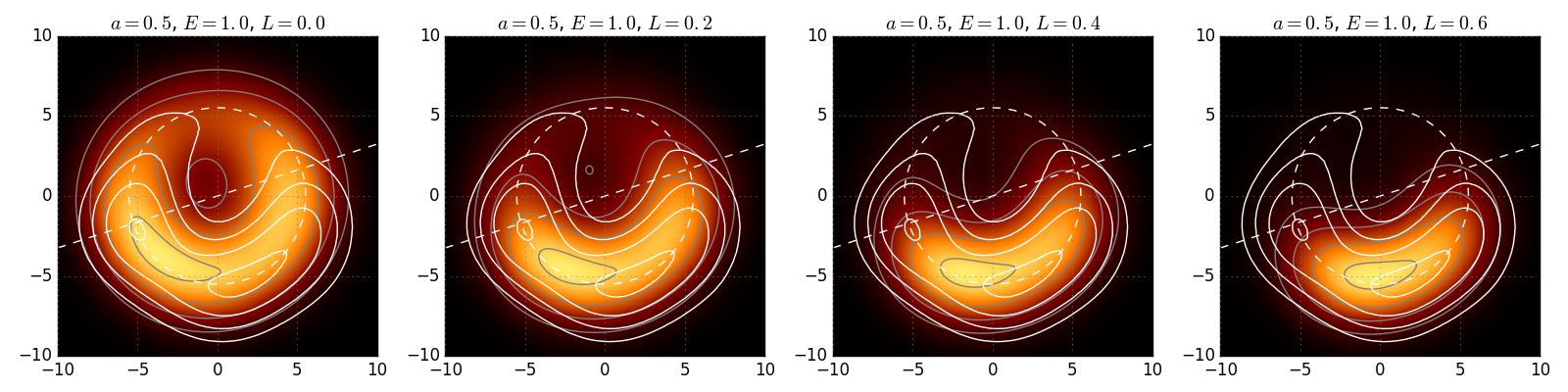}
\caption{The case of polar caps (TH1) for $a = 0.5$ with fixed values of conserved energy $E$ and angular momentum $L$. See Fig. \ref{fig_eht_beta0_spin_theta} for more description.}
\label{fig_eht_caps_EL}
\end{figure*}

\begin{figure}
\centering
\includegraphics[width=\columnwidth]{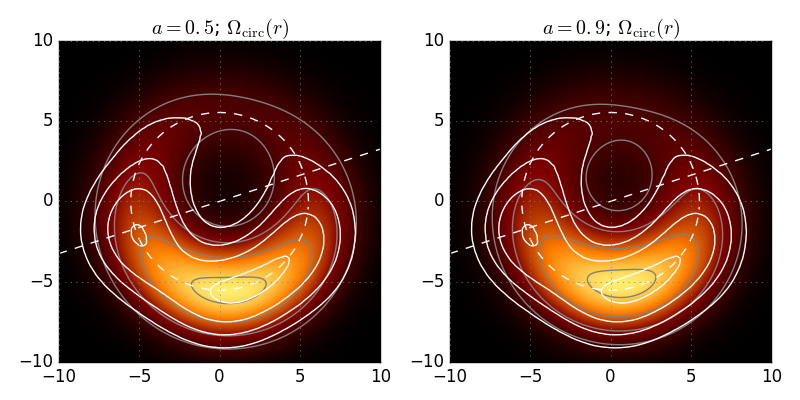}
\includegraphics[width=\columnwidth]{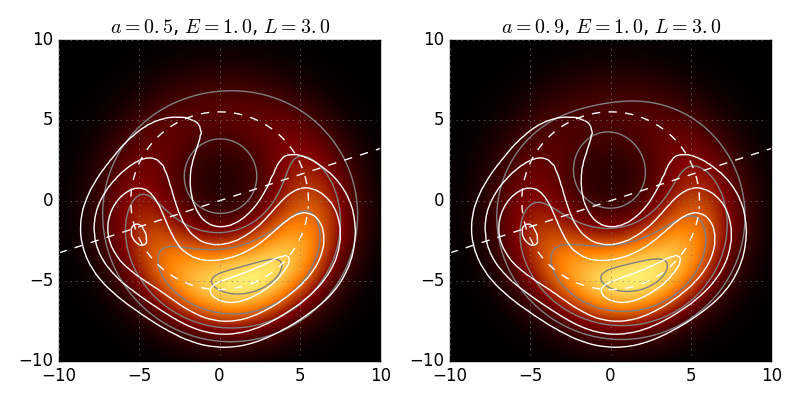}
\caption{The case of equatorial ring (TH4) for two values of BH spin, with different assumptions on the kinematics of the emitting fluid. \emph{Top panels}: stable circular orbits with $\beta_r = 0$ and Keplerian angular velocity $\Omega_{\rm circ}(r)$ within prograde discs limited by $r_{\rm ISCO}(a) < r < r_{\rm max} = 6M$. \emph{Bottom panels}: plunging orbits with conserved energy $E$ and angular momentum $L$, limited by $r_{\rm h} < r < r_{\rm max} = 6M$. See Fig. \ref{fig_eht_beta0_spin_theta} for more description.}
\label{fig_eht_kepler_EL}
\end{figure}

\begin{figure*}
\centering
\includegraphics[width=\textwidth]{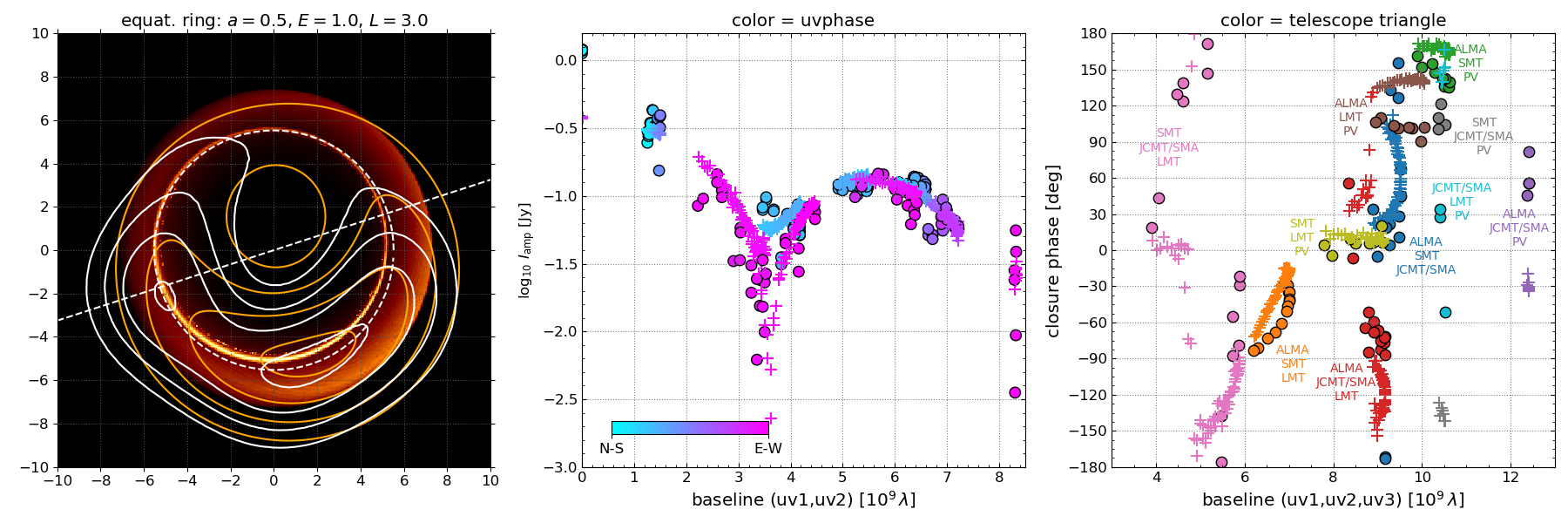}
\includegraphics[width=\textwidth]{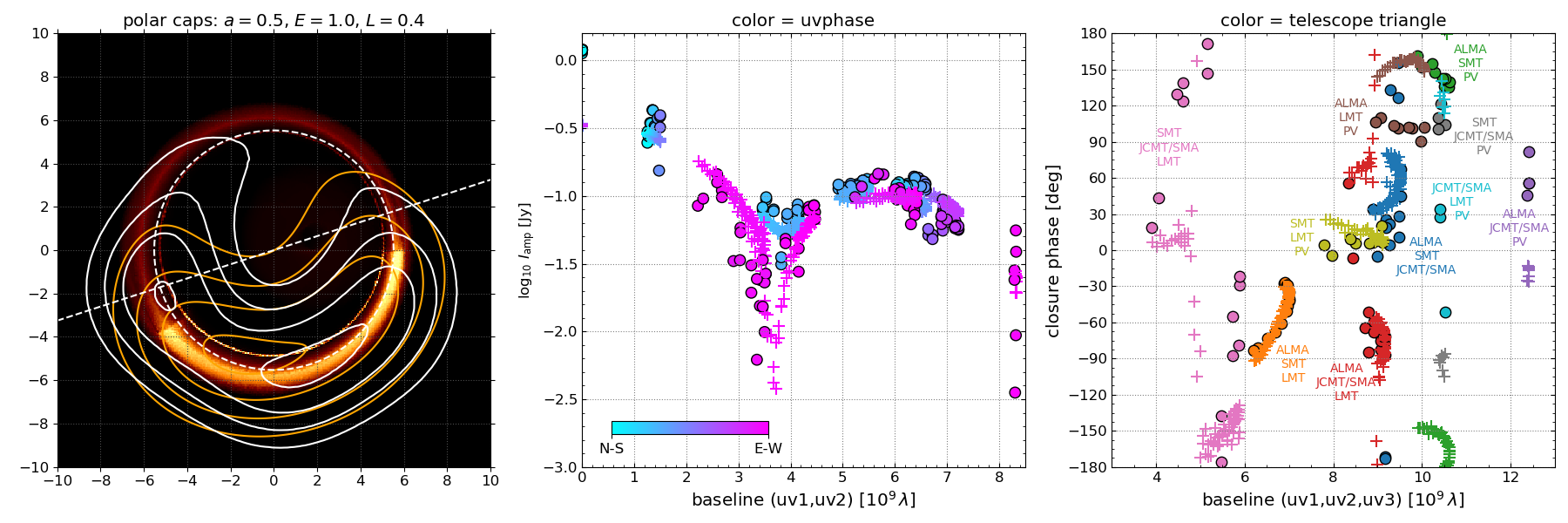}
\caption{Results of simulated EHT observations of selected model images.
\emph{The left panels} show the model images (high resolution with the colour scale and EHT resolution with the orange contours) compared with the reference image reconstructed from the actual EHT observations of M87* obtained on 2017 April 11th (solid white contours).
\emph{The middle panels} show the visibility amplitude as function of telescope pair baseline. The filled circles indicate the actual EHT measurements of M87*, and the crosses indicate the simulated EHT observations of the model image. The symbol colour indicates the orientation of the telescope pair baseline.
\emph{The right panels} show the closure phase as function of telescope triangle baseline. The symbol colour indicates the telescope triangle indicated by the labels.}
\label{fig_closures}
\end{figure*}

\section{Results}

Fig. \ref{fig_eht_calib} shows our rendering of the EHT image of M87* {for 2017 April 11, adapted from Fig. 3 in \cite{2019ApJ...875L...1E},} that we use for reference. The peak flux density in terms of brightness temperature is $5.7\times 10^9\;{\rm K}$.

We first study the effect of location of the emitting regions in the $\theta$ space, considering the following four cases:
(TH1) $\theta_{\rm min} = 0$ and $\theta_{\rm max} = 30^\circ$ (polar caps);
(TH2) $\theta_{\rm min} = 30^\circ$ and $\theta_{\rm max} = 45^\circ$;
(TH3) $\theta_{\rm min} = 45^\circ$ and $\theta_{\rm max} = 75^\circ$;
(TH4) $\theta_{\rm min} = 75^\circ$ and $\theta_{\rm max} = 90^\circ$ (equatorial ring).
The results are shown in Fig. \ref{fig_eht_beta0_spin_theta} for two extreme values of BH spin: $a = 0$ and $a = 0.9$. We note that there is only a mild effect of BH spin on the images produced with $r_{\rm max} = 6M$.

In the case of polar caps (TH1), the image is dominated by a single compact hot spot slightly offset from the image centre along the jet direction, coinciding with the middle of the shadow observed by the EHT. This is an image of the front polar cap, while the image of the back polar cap is smeared roughly uniformly along the photon ring.
In the case TH2, the image of the front cap becomes extended, but remains centrally peaked.
In the case TH3, the image morphology turns into a ring of radius $\sim 4 R_{\rm g}$ without a deep shadow on top of a roughly uniform image of the back ring spanning the entire photon ring.
Finally, in the case of equatorial ring (TH4), the size of the simulated image becomes consistent with the size of the EHT image, but the brightness distribution along the ring is fairly uniform. The effect of BH spin is that for $a = 0$ the brightest sector of the ring is that opposite to the jet direction, while for $a = 0.9$ it is rotated towards the southern side (CW in the case of TH2 and CCW in the case of TH3). Without additional Doppler beaming, there is too much emission from the northern part of the ring.

We now consider the effect of Doppler beaming due to the motion of the emitting plasma. First, in the case of polar caps (TH1), we study the effect of radial infall with fixed coordinate velocity $\beta_r$.
Fig. \ref{fig_eht_th1_betar} shows that the image of the front cap is strongly suppressed for the radial velocity of $\beta_r \ge 0.4$, and we see the image of the back cap magnified into a ring that is slightly brighter on the southern side. This illustrates the basic fact that kinematics of emitting fluid, which involves mildly relativistic motions, has a very strong effect on the appearance of BH environments.

However, the case of fixed radial velocity, including that of static emitter fluid, is physically unrealistic. Hence, as the next level of approximation, we consider plunging ($\beta_r < 0$) timelike geodesic flows that conserve energy $E$ and angular momentum $L$. We neglect the poloidal 4-velocity component $u^\theta = 0$. Since the 4-velocity components $u^t$ and $u^\phi$ can be calculated explicitly from $E$ and $L$, we can then calculate the radial component $u^r$ from $u^2 = -1$. We consider the fluid to be constricted to regions where $(u^r)^2 \ge 0$.

Fig. \ref{fig_eht_caps_EL} presents the case of polar caps (TH1) for the BH spin value of $a = 0.5$ and fixed asymptotic energy $E = 1$, showing the effect of moderate angular momentum $0 \le L \le 0.6$. The case of $L = 0$ is very similar to the case of $\beta_r = -0.4$ -- it is a fairly uniform ring with maximum brightness in the SE sector. Introducing non-zero $L$ has a very strong effect for decreasing the brightness along the northern side of the ring, turning it into a crescent already for $L = 0.2$, although the N-S contrast is still too low, and the second-brightest contour of the crescent is rotated CCW with respect to the observed image. Higher values of $L$ make the image too compact, centred in the S sector.

In the case of equatorial rings (TH4), we consider either the plunging geodesic timelike flows with conserved $E$ and $L$, or quasi-Keplerian circular motions with $\beta_r = 0$ and angular velocity $\Omega = \Omega_{\rm circ}$, with the rings limited from the inside by the ISCO (innermost stable circular orbit; \citealt{1972ApJ...178..347B}) and from the outside by $r_{\rm max} = 6M$ (we only consider prograde disks for $a > 0$, for which $R_{\rm ISCO} < 6M$).

Fig. \ref{fig_eht_kepler_EL} compares the images calculated in these two scenarios for two values of BH spin $a = 0.5, 0.9$. It is notable that these images are very similar to each other, with most of the emission concentrated along the southern side of the photon ring. Our models for plunging flows with fixed $E$ and $L$ provide very natural match with the observed emission in the SSE-SWW sector. On the other hand, emission from the SEE sector is very weak in all cases. It should be stressed that these results are not very sensitive to the choice of $E$ and $L$ values, since the images are very similar for $0.94 \le E \le 1.3$ and for $2 \le L \le 3.9$. Most of these values of $E$ and $L$ allow for flows plunging onto the BH horizon.

\subsection{Simulated EHT observations}

Figure \ref{fig_closures} shows the results of simulated EHT observations of two representative images. We present the visibility amplitudes $I_{\rm amp,ij}$ as functions of pair baselines $(u_{ij}^2 + v_{ij}^2)^{1/2}$, and the closure phases $\Psi_{C,ijk}$ as functions of triangle baselines $(u_{ij}^2 + u_{ik}^2 + u_{jk}^2 + v_{ij}^2 + v_{ik}^2 + v_{jk}^2)^{1/2}$, and we compare them with the results of EHT observation of M87 obtained on 2017 April 11th in the high frequency band.

The first simulated image is the case of an equatorial ring with parameters $a = 0.5$, $E = 1$ and $L = 3$.
We obtain a reasonable agreement between the simulated and observed visibility amplitudes.
We note that the normalisation of simulated $I_{\rm amp}$ values is approximate, it is intended to match the observed visibility for baselines in the range $\sim (5-7)\;{\rm G\lambda}$, however, our model underproduces the total observed flux ($I_{\rm amp}$ in the limit of zero baseline) by factor $\sim 3$.
The first minimum of the simulated $I_{\rm amp}$ is obtained for the E--W baselines of $\simeq 3.6\;{\rm G\lambda}$, slightly longer than for the observed minimum at $\simeq 3.4\;{\rm G\lambda}$.
This minimum is much shallower for the N--S baselines, both for the simulated and real observations.\footnote{This is entirely consistent with the results presented by the EHT Collaboration in the Figure 1 of \cite{2019ApJ...875L...6E}, where the E--W baselines are shown in blue and the N--S baselines are shown in red.}
The second minimum of $I_{\rm amp}$ at the E--W baselines $\simeq 8.3\;{\rm G\lambda}$ is not very deep in the simulated model data.

The closure phases reveal various levels of agreement between the simulated observations of our model and actual measurements of M87*.
A reasonable agreement is found for the SMT-LMT-PV and ALMA-SMT-LMT triangles characterised by intermediate values of triangle baselines $\simeq (6-9)\;{\rm G\lambda}$, as well as for the ALMA-SMT-PV triangle with at $(10-10.5)\;{\rm G\lambda}$.
On the other hand, there is a clear discrepancy at the shortest triangle baselines $\simeq (3.9-4.6)\;{\rm G\lambda}$, where the simulated $\Psi_{C,\rm SMT-JCMT/SMA-LMT}$ is consistent with zero, while its observed values span a wide range of intermediate values $(20^\circ:140^\circ)$.
There are also systematic differences between closure phases for the ALMA-LMT-PV ($\simeq 140^\circ$ simulated vs. $\simeq 100^\circ$ observed), ALMA-JCMT/SMA-LMT (mostly $[-150^\circ:-90^\circ]$ simulated vs. $[-90^\circ:-50^\circ]$ observed) and ALMA-SMT-JCMT/SMA (mostly $[20^\circ:110^\circ]$ simulated vs. $[-10^\circ:70^\circ]$ observed) triangles.

The second simulated image is the case of a polar cap with parameters $a = 0.5$, $E = 1$ and $L = 0.4$.
The structure of the visibility amplitude as function of the pair baseline is very similar to the previous case.
The first minimum of $I_{\rm amp}$ is found at the slightly longer E--W pair baseline of $3.7\;{\rm G\lambda}$.
The simulated closure phases are generally similar to the previous case.
Much better agreement is found for the triangles ALMA-SMT-LMT and ALMA-JCMT/SMA-LMT (main group).
On the other hand, the simulated values for the triangle ALMA-LMT-PV are shifted to $\simeq 155^\circ$, and those for the triangle ALMA-SMT-PV are found at $[-180^\circ:-150^\circ]$.
For the smallest triangle SMT-JCMT/SMA-LMT, the closure values measured for triangle baselines longer than $5\;{\rm G\lambda}$ are concentrated at $[-170^\circ:-130^\circ]$.

\section{Discussion and conclusions}

The EHT image of M87* BH could be interpreted either as a single crescent extending from the E to SWW sectors of the photon ring, or as a combination of a short crescent in the SSE-SWW sector and a compact SEE `hotspot'. The GRMHD simulations of geometrically thick accretion onto BHs tend to produce images of crescents that are roughly parallel to the projected jet axis \citep{2012MNRAS.421.1517D,2016A&A...586A..38M,2019A&A...632A...2D}. Hence, these simulations can explain emission from the SSE-SWW sector as naturally as our simplified toy models.
This qualitative consistency of results suggests that the SEE emission might require a distinct origin.

In our models of polar caps (TH1), the image of the front polar cap is expected to coincide with the observed position of the BH shadow. In Fig. \ref{fig_eht_th1_betar}, we demonstrate that radial infall with velocity of $\beta_r \sim 0.4$ is able to suppress the image of the front cap, at the same time enhancing the extended image of the back cap.
In turn, Fig. \ref{fig_eht_caps_EL} shows that the image of the back cap is sensitive to the angular momentum of the plunging flow, with $L \simeq 0.4$ it is possible to reproduce the required north-south contrast, however, emission from the SEE sector is insufficient to explain the EHT observations.

Our models of equatorial rings (TH4) with $r_{\rm max} \simeq 6 R_{\rm g}$ (Fig. \ref{fig_eht_kepler_EL})
produce crescent images that typically extend from SSE to W position angles, with only weak emission in the SEE sector,
almost independently of the adopted kinematic model for the emitter fluid.
We considered stable equatorial orbits limited from the inside by ISCO (prograde orbits for $a > 0$) and plunging flows with conserved energy and angular momentum, and in no case were we able to extend the crescent towards the SEE sector.

The differences between our model images and the reconstructed EHT image of M87* are reflected in the differences between the closure phases obtained by simulating EHT observations of our images and those actually measured during the EHT observations of M87*.
In our models, we can reproduce the closure phases for certain telescope triangles (e.g. ALMA-SMT-LMT), but never for all of them, in particular it is difficult to reproduce the highly variable closure phases for the smallest triangle SMT-JCMT/SMA-LMT.

The EHT image of M87* does not allow to significantly constrain the value of BH spin due to the fact that the radius of the photon ring is not sensitive to the spin value \citep{2019ApJ...875L...5E}, and the photon ring departs substantially from circularity only for $a \ge 0.95$ \citep{2019PhRvD.100d4057B}.
Stronger constraints can be put from the requirement for production of sufficiently powerful jets, which essentially eliminates the values $|a| < 0.4$ \citep{2019MNRAS.489.1197N,2019ApJ...880L..26N}.
We find that considering extreme spin values of $a = 0$ and $a = 0.9$ does not qualitatively affect the obtained images for $r_{\rm max} = 6 R_{\rm g}$. This effect would be much greater for smaller values of $r_{\rm max}$ (e.g., 3), which would however result in significantly smaller images (for the same value of BH mass).
On the other hand, {\cite{2019Univ....5..183D} argued} that the relative orientation of the BH shadow and the brightness centre of the crescent constraints the spin value to $a = 0.75\pm 0.15$. That result was obtained by ray tracing emission from plunging geodesic flows with conserved $E$ and $L$ in an optically thin equatorial disk limited to \emph{within} the ISCO radius. That study suggests that the SEE `hotspot' coincides with the brightest point of the accretion flow, oriented roughly perpendicularly to the BH spin, which is however not consistent with the direction of the large-scale jet.

Our models were calculated under strict assumptions of azimuthal and planar symmetry of the emitting regions. The easiest way to obtain a crescent image of correct orientation is to relax the assumption of azimuthal symmetry, adding an $m=1$ mode to the local emissivity. However, from such relaxed models we cannot draw any meaningful constraints on the geometry and velocity fields of the underlying plasma flows. In particular, emission in the SEE sector could be explained by a localised and temporary perturbation in the accretion flow. One possibility for such perturbations are filaments resulting from the interchange instability operating in the magnetically arrested disks (MAD) \citep{2012MNRAS.423.3083M}.
Any such temporary effects should not persist in the subsequent EHT observations.
Short-term variations of the EHT images of M87* are at the level of $\lesssim 15\%$ over the time range of 6 days, equivalent to $16R_{\rm g}/c$ \citep[Fig. 33]{2019ApJ...875L...4E}.

It has been suggested that synchrotron self-absorption may have a significant effect on the image of M87* at the $1.3\;{\rm mm}$ wavelength if observed eventually in a higher flux state \citep{2019ApJ...878...27K}.
A potentially more important factor could be anisotropic rest-frame emissivity due to the synchrotron process operating in ordered magnetic fields.
While GRMHD simulations are likely to produce realistic distributions of magnetic fields, they typically treat the emitting particles as locally isotropic.
On the other hand, it has been argued in the context of relativistic AGN jets that the momenta of energetic electrons could be strongly concentrated along the local magnetic field lines \citep[e.g.,][]{2019MNRAS.484.1192S}.
Global kinetic numerical simulations or improved subgrid MHD models may be required to incorporate such effects.

The origin of emission observed by the EHT in the SEE sector of the photon ring in M87* cannot be explained using strictly stationary and axisymmetric models -- whether equatorial rings or polar caps -- that assume the theoretically expected alignment of the BH spin with the large-scale jet.
The most natural solution is to invoke temporary fluctuations of the inner accretion flow magnified along the photon ring \citep{2019ApJ...875L...5E}.
The SEE `hotspot' can thus be expected to disappear in the subsequent EHT observations of M87*.

\begin{acknowledgement}

We thank the anonymous reviewer for helpful suggestions.
We acknowledge discussions with Bartosz Be{\l}dycki, Beno{\^i}t Cerutti, Christian Fromm, Jean-Pierre Lasota and Maciej Wielgus.
We have used the {\tt eht-imaging} software \citep{2016ApJ...829...11C,2018ApJ...857...23C} obtained from \url{https://github.com/achael/eht-imaging}, and the EHT Science Release 1 (SR1) dataset \citep{SR1} obtained from \url{https://eventhorizontelescope.org/for-astronomers/data}.
This work was partially supported by the Polish National Science Centre grants 2015/18/E/ST9/00580 and 2015/17/B/ST9/03422.
\end{acknowledgement}

% REFERENCES

\end{document}